\documentclass[ 
onecolumn,
prl, tightenlines,
nofootinbib,
amsfonts,amsmath,amssymb]{revtex4}

\usepackage{bm} \usepackage{graphics,graphicx,epsfig,color}
\usepackage{subfigure} \usepackage{hyperref}

\begin{document}
 
\title{MOND avec ou sans mati\`ere noire\,\footnote{Version plus d\'etaill\'ee d'un article paru dans\,: \textit{La Recherche}, no. {\bf 435}, p. 44 (novembre 2009).}}

\author{Luc \textsc{Blanchet}} 
\affiliation{Institut d'Astrophysique de Paris, GRECO, 98$^\text{bis}$ boulevard Arago, 75014 Paris, France}
\author{Fran\c{c}oise \textsc{Combes}} 
\affiliation{Observatoire de Paris, LERMA, 61 avenue de l'Observatoire, 75014 Paris, France}


\begin{abstract}
Dans le mod\`ele cosmologique, l'Univers est constitu\'e principalement de mati\`ere noire et d'\'energie noire dont nous ne connaissons pas la nature, et pour lesquelles il n'y a pas d'explication dans le cadre du mod\`ele standard de la physique des particules. Comment incorporer mati\`ere et \'energie noires dans l'ensemble des lois fondamentales\,? Toutes les observations peuvent aussi bien \^etre expliqu\'ees soit par l'addition de composants de l'Univers inconnus, avec la relativit\'e g\'en\'erale comme th\'eorie de la gravitation, soit par une modification fondamentale de cette th\'eorie. Ne serait-il pas plus simple de la modifier\,? C'est le cas de l'hypoth\`ese MOND (pour MOdified Newtonian Dynamics) propos\'ee par Milgrom en 1983, qui est pleine de succ\`es pour d\'ecrire la cin\'ematique et la dynamique des galaxies. Il serait cependant possible d'obtenir le m\^eme succ\`es gr\^ace \`a une nouvelle forme de mati\`ere, la mati\`ere noire dipolaire, tout en gardant la relativit\'e g\'en\'erale comme loi de la gravitation.
\end{abstract}


\maketitle

Durant la derni\`ere d\'ecennie, la cosmologie observationnelle a fait d'\'enormes progr\`es\,: \`a l'aube de l'an 2000, nous ne connaissions ni l'\^age de l'univers, ni sa g\'eom\'etrie, ni son contenu avec la moindre pr\'ecision. La constante de Hubble, qui mesure l'expansion de l'Univers \`a partir du Big-Bang, \'etait encore incertaine, pouvant varier d'un facteur 2. Aujourd'hui, tous les scientifiques reconnaissent que la cosmologie est devenue une science de pr\'ecision. Gr\^ace \`a l'\'etude d\'etaill\'ee du fond de rayonnement cosmologique, vestige du Big-Bang, et de ses irr\'egularit\'es, gr\^ace \`a l'observation pr\'ecise de supernov{\ae} comme indicateurs de distance, et gr\^ace \`a l'observation de la d\'eviation des rayons lumineux par la mati\`ere, nous avons pu d\'eterminer au moins \`a 10\% pr\`es tous les param\`etres de l'Univers, \^age, courbure, et composition, cf. la Figure 1. Une conclusion fascinante est que la mati\`ere ordinaire ne compte que 4\% dans ce recensement, et m\^eme une grande partie de la mati\`ere ordinaire sous forme d'atomes qui ne rayonnent pas n'a pas \'et\'e identifi\'ee.
\begin{figure}[h]
\centering{\includegraphics[width=3.5in]{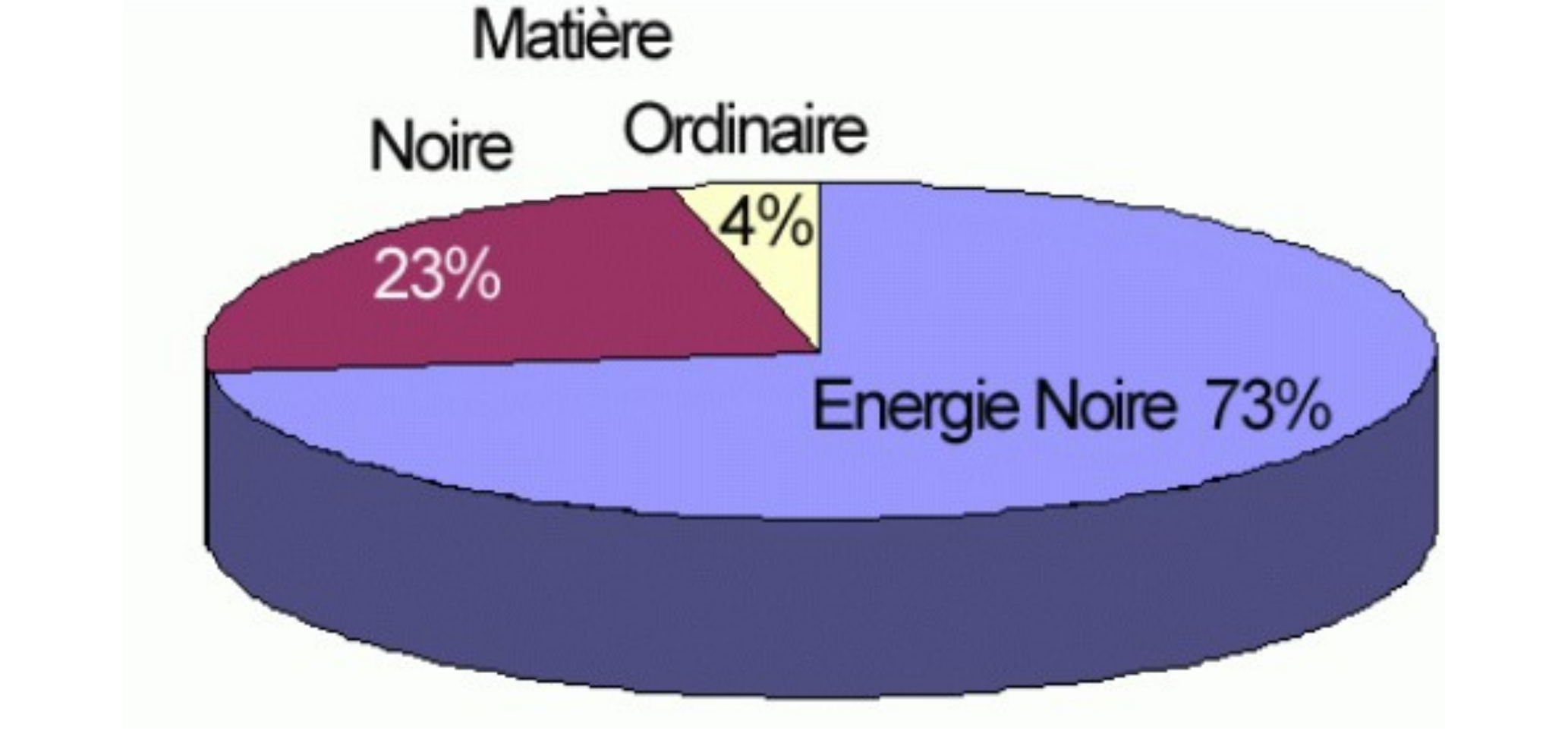}
\caption{Contenu de l'Univers \`a grande \'echelle.}
\label{fig1}}
\end{figure}

\section{Deux nuages sombres \`a l'horizon}

Les avanc\'ees consid\'erables de ces derni\`eres ann\'ees mettent donc en \'evidence une grande tension dans notre vision de l'Univers\,: celui-ci est constitu\'e essentiellement de mati\`ere noire et d'\'energie noire dont nous ne connaissons pas la nature, et pour lesquelles il n'y a pas d'explication dans le cadre actuel de la physique des particules. La mati\`ere noire, de nature exotique, pourrait \^etre form\'ee de particules pr\'edites par certaines extensions du mod\`ele standard comme dans les th\'eories supersym\'etriques. Mais ces particules n'ont toujours pas \'et\'e mises en \'evidence de mani\`ere directe, malgr\'e de nombreuses exp\'eriences ayant tent\'e de les d\'etecter (peut-\^etre que le LHC va bient\^ot nous en dire plus). Quant \`a l'\'energie noire, c'est un nouveau composant myst\'erieux qui acc\'el\`ere l'expansion de l'Univers, et qui pourrait \^etre la fameuse constante cosmologique $\Lambda$ qu'Einstein avait introduite dans les \'equations de la relativit\'e g\'en\'erale, pour des raisons maintenant disparues. Le probl\`eme est que toutes les tentatives pour interpr\'eter $\Lambda$ en termes de physique fondamentale ont pour l'instant \'echou\'e.

Mati\`ere noire et \'energie noire sont comme deux nuages qui se profilent \`a l'horizon du ciel radieux de la physique. Sommes-nous \`a l'aube d'une r\'evolution fondamentale\,? Plusieurs satellites, comme Euclid en Europe, ou JDEM aux \'Etats-Unis, vont mesurer plus pr\'ecis\'ement les caract\'eristiques des composantes noires, mais on peut argumenter que le probl\`eme est plut\^ot d'ordre th\'eorique\,: comment incorporer mati\`ere et \'energie noires dans un ensemble de lois fondamentales\,?

Toutes les observations peuvent aussi bien \^etre expliqu\'ees soit par l'addition de composants inconnus, en restant dans le cadre de la relativit\'e g\'en\'erale comme th\'eorie de la gravitation, soit par une modification fondamentale de cette th\'eorie. Ne serait-ce pas plus simple de la modifier\,? Au XIX$^\text{\`eme}$ si\`ecle l'astronome fran\c{c}ais Le Verrier avait d\'ecouvert une rotation anormale (dite pr\'ecession) de l'orbite de la plan\`ete Mercure. On pouvait l'expliquer soit par de la ``mati\`ere noire'', en l'occurrence une nouvelle plan\`ete int\'erieure \`a l'orbite de Mercure (d\'enomm\'ee Vulcain par Le Verrier), soit par une modification des lois de Newton. C'est cette deuxi\`eme option qui s'est finalement av\'er\'ee correcte, lorsque la gravitation newtonienne a \'et\'e remplac\'ee par celle d'Einstein, et que l'exc\`es de pr\'ecession de Mercure a \'et\'e expliqu\'e par un effet relativiste.
\begin{figure}[t]
\centering{\includegraphics[width=4in]{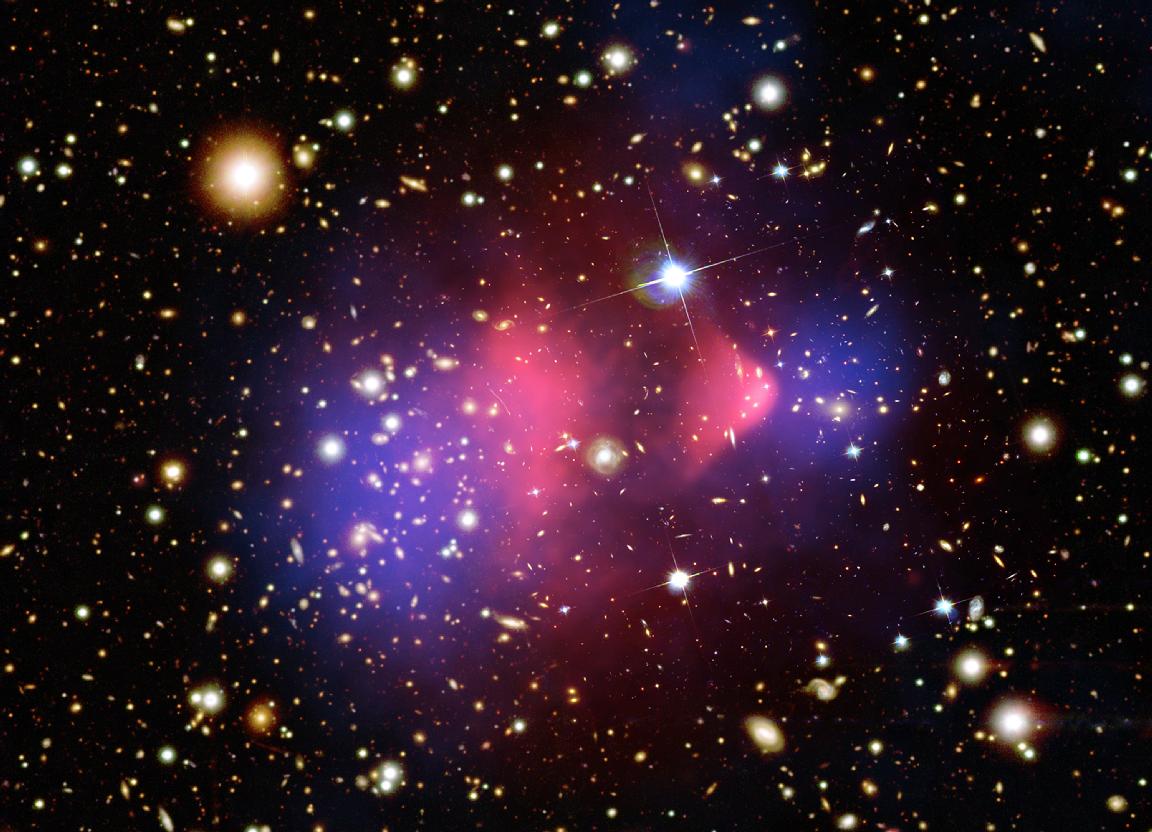}
\caption{L'amas de galaxies dit ``du boulet''. Il s'agit en fait de deux amas de galaxies en collision. \`A la photo optique du syst\`eme (qui montre les deux amas de galaxies comme une concentration de t\^aches blanches) est superpos\'ee en rouge l'\'emission X du gaz chaud, et en bleu la masse totale projet\'ee. La distribution de masse projet\'ee sur le ciel correspond \`a la masse des amas reconstruite par lentilles gravitationnelles. Le plus petit amas de galaxies \`a droite semble avoir travers\'e comme un boulet le gros amas \`a gauche. La pointe de couleur rouge \`a droite montre clairement une onde de choc en forme de c\^one, qui permet de mesurer la vitesse supersonique de cette collision (ici 4700 km/s, ou Mach 3). Tous les composants de l'amas ne r\'eagissent pas de la m\^eme fa\c{c}on dans la collision\,: les galaxies et la mati\`ere noire peuvent s'interp\'en\'etrer sans presque se voir. En revanche, le gaz chaud est frein\'e, si bien que les deux composants gazeux des deux amas sont plus rapproch\'es que les deux masses totales. Le comportement diff\'erent dans la collision du gaz chaud, des galaxies et de la mati\`ere noire permet de les s\'eparer, et de tester les mod\`eles. D'apr\`es Clowe et al. (2006).}
\label{fig2}}
\end{figure}

\section{Probl\`eme de la masse manquante}

La mati\`ere noire est un composant connu depuis les ann\'ees 1930, quand l'astronome suisse F. Zwicky avait besoin de 100 fois plus de masse que la mati\`ere visible, pour expliquer la dynamique des amas de galaxies. Depuis, nous avons d\'ecouvert le gaz tr\`es chaud, \'emetteur de rayons X, qui repr\'esente l'essentiel de la mati\`ere visible dans les amas, mais la mati\`ere noire est encore dominante, et \'egale \`a environ 6 fois la mati\`ere visible dans ces environnements. Dans les ann\'ees 1980, les courbes de rotation des galaxies spirales, tr\`es bien \'etablies par les observations, ont aussi mis en \'evidence la mati\`ere manquante au niveau des galaxies, et depuis une dizaine d'ann\'ees, l'outil des lentilles gravitationnelles permet de cartographier la mati\`ere noire, au voisinage des grandes structures\,: les galaxies de fond envoient des rayons lumineux qui sont d\'evi\'es par la mati\`ere sur la ligne de vis\'ee avant d'arriver \`a l'observateur. La d\'eformation des images, due \`a ces ``lentilles'', est mesur\'ee et trait\'ee statistiquement pour reproduire la distribution de mati\`ere totale, comme par exemple sur la Figure 2.

La mati\`ere noire est aussi corrobor\'ee par les pics de fluctuations du fond de rayonnement cosmologique, ou rayonnement de corps noir \`a 2,7 degr\'es Kelvin, \'emis peu de temps apr\`es le Big-Bang. La position et la hauteur des pics indiquent la pr\'esence de mati\`ere noire sous forme d'un fluide de particules froides (``Cold Dark Matter'' ou CDM) sans interactions. 

Le mod\`ele cosmologique actuel $\Lambda$-CDM rencontre beaucoup de succ\`es \`a grande \'echelle, comme le montrent les simulations num\'eriques de la formation des grandes structures (cf. Figure 3). Dans ce mod\`ele la mati\`ere noire joue un r\^ole crucial car c'est elle qui entra\^ine la mati\`ere ordinaire dans un effondrement gravitationnel et permet d'expliquer la formation des structures.
\begin{figure}[t]
\centering{\includegraphics[width=4in]{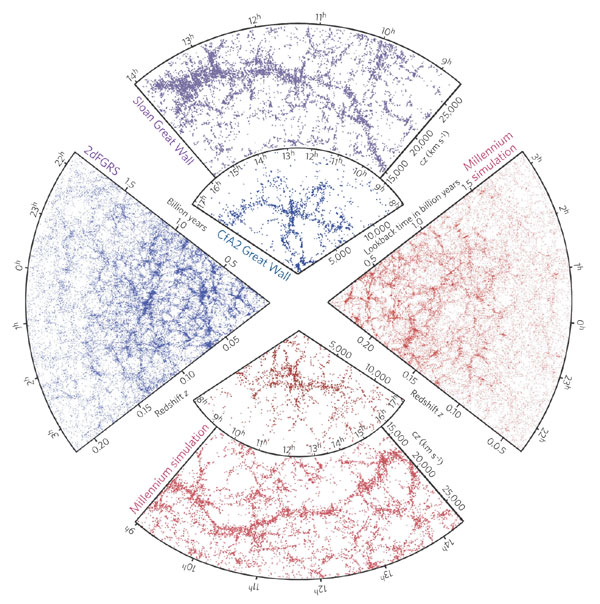}
\caption{Comparaison entre les cartographies de galaxies \`a grande \'echelle (en bleu), et les simulations num\'eriques (en rouge). En haut et \`a gauche sont montr\'ees des tranches d'univers, o\`u chaque point est une galaxie, dont la distance est obtenue par le d\'ecalage vers le rouge, mesur\'e sur son spectre (catalogues SDSS et 2dF). En bas et \`a droite, pr\'esent\'ees de la m\^eme fa\c{c}on, les pr\'edictions des simulations de mati\`ere noire Millenium. D'apr\`es Springel et al. (2006).}
\label{fig3}}
\end{figure} 

\section{Probl\`emes de CDM \`a l'\'echelle des galaxies}

Si le mod\`ele CDM est correct il doit pouvoir aussi expliquer les halos observ\'es de mati\`ere noire autour des galaxies. Malheureusement, les pr\'edictions des simulations \`a l'\'echelle des galaxies posent de nombreux probl\`emes. 

La mati\`ere noire CDM se concentre beaucoup trop dans les galaxies, et une spirale comme la Voie Lact\'ee par exemple devrait \^etre domin\'ee par la mati\`ere noire, m\^eme dans sa partie centrale, ce qui n'est pas observ\'e. D'autre part, les simulations sugg\`erent un profil de densit\'e de mati\`ere noire autour des galaxies o\`u la distribution radiale de mati\`ere noire doit monter tr\`es vite vers le centre, et y former un pic de densit\'e, alors que les courbes de rotation indiquent plut\^ot la pr\'esence de plateaux de densit\'e constante au centre. 

La formation de galaxies s'effectue en grande partie par fusions de galaxies plus petites, dans le sc\'enario dit ``hi\'erarchique''. Les fusions entre galaxies sont acc\'el\'er\'ees par la perte d'\'energie orbitale au profit des halos de mati\`ere noire. La mati\`ere ordinaire spirale tr\`es vite vers le centre des halos, en perdant une grande partie de sa rotation. Ainsi, dans le mod\`ele standard, la taille des disques simul\'es est pr\`es de dix fois plus petite par rapport \`a la taille des galaxies spirales observ\'ees. De plus, comme le montre la Figure 4, le sc\'enario pr\'edit un grand nombre de satellites autour d'une galaxie g\'eante typique, comme la Voie Lact\'ee. Ces compagnons ne sont pas observ\'es, et devraient donc avoir perdu toutes leurs \'etoiles et leur gaz.
\begin{figure}[t]
\centering{\includegraphics[width=4in]{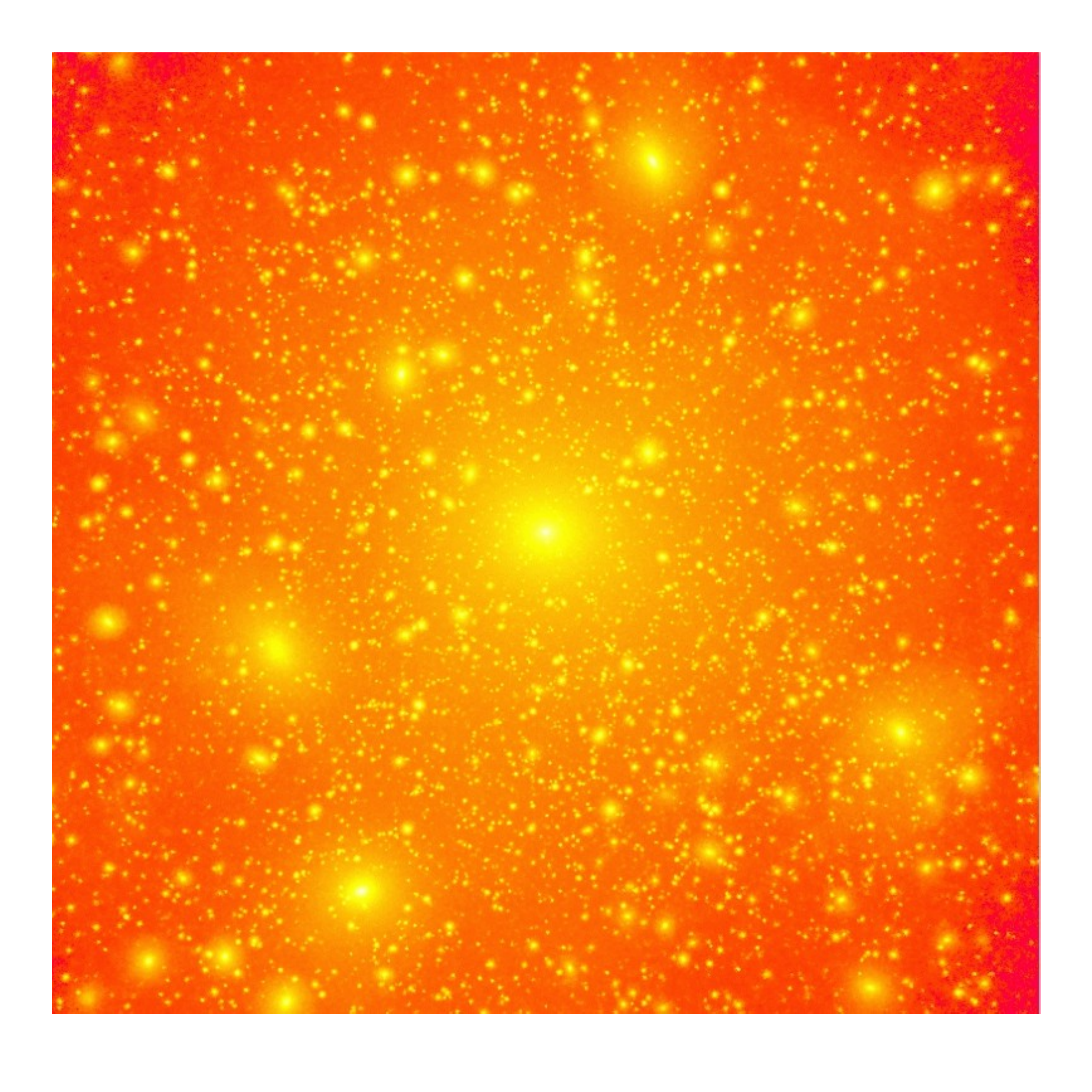}
\caption{Simulation de la formation de la Voie Lact\'ee, dans le mod\`ele standard de mati\`ere noire CDM\,: plus d'un millier de satellites devrait orbiter autour du centre, alors que nous en observons une douzaine. D'apr\`es Diemand et al. (2007).}
\label{fig4}}
\end{figure} 

Un autre probl\`eme, qui semble de plus en plus fondamental, est l'\'evidence observationnelle de l'existence d'une nouvelle constante ``universelle'' d'acc\'el\'eration $a_0$ mesur\'ee \`a la valeur $a_0=1,2\,10^{-10}\,\mathrm{m}/\mathrm{s}^2$. Selon la loi de l'isra\'elien Moti Milgrom, la mati\`ere noire ne se manifeste que dans les r\'egions o\`u le champ de gravitation est plus faible que $a_0$. Cette loi \'etrange mais tr\`es bien v\'erifi\'ee est inexplicable dans le cadre CDM. 

Plus g\'en\'eralement, on observe une surprenante et myst\'erieuse r\'egularit\'e dans la distribution de mati\`ere noire autour des galaxies, qui se traduit par une sorte de ``conspiration'' entre la mati\`ere noire et la mati\`ere visible pour rendre compte des observations. Dans le mod\`ele CDM, il faut ajuster finement la proportion et la distribution de mati\`ere noire dans chaque galaxie, alors que l'on s'attendrait \`a une dispersion plus grande des propri\'et\'es des halos de mati\`ere noire, \`a cause du hasard de l'histoire de chaque galaxie et de son environnement. 

Ne sommes-nous pas en train de faire une erreur en extrapolant le mod\`ele standard de l'\'echelle cosmologique \`a l'\'echelle galactique\,?

\section{La gravit\'e newtonienne modifi\'ee (MOND)}

Cette ``r\'egularit\'e'' des halos galactiques s'exprime par une formule purement empirique qui permet d'ajuster avec grande pr\'ecision les courbes de rotation des galaxies, et de reproduire la c\'el\`ebre relation des astronomes am\'ericains Brent Tully et Richard Fisher, qui relie la vitesse de rotation des galaxies et leur luminosit\'e (celle-ci varie comme la puissance 4$^\text{\`eme}$ de la vitesse), et qui est mal comprise dans le mod\`ele CDM.

Dans une r\'egion o\`u le champ de gravitation est plus faible que la constante $a_0$, tout se passe comme si le champ changeait de r\'egime, passant de la valeur newtonienne $g_\mathrm{N}$, qui d\'ecro\^it comme $1/r^2$ loin de la galaxie, \`a la valeur $g=\sqrt{g_\mathrm{N}\,a_0}$. Donc, selon cette formule, $g$ va d\'ecro\^itre loin de la galaxie en $1/r$, soit moins rapidement que le champ newtonien, et on ``explique'' pourquoi les \'etoiles tournent plus vite autour de la galaxie sans avoir recours \`a la mati\`ere noire. Celle-ci ne serait donc qu'une apparence, et c'est la loi de la gravitation qui serait modifi\'ee. C'est l'hypoth\`ese MOND (pour MOdified Newtonian Dynamics) propos\'ee par M. Milgrom en 1983. 

Dans le sc\'enario MOND, les courbes de rotation sont remarquablement reproduites pour tous les types de galaxies, des g\'eantes peu concern\'ees par le probl\`eme de la masse manquante, jusqu'aux galaxies naines, enti\`erement domin\'ees par la mati\`ere noire. 

Des simulations num\'eriques de la dynamique des galaxies dans le cadre du formalisme de MOND ont \'et\'e effectu\'ees par O. Tiret et F. Combes dans les trois derni\`eres ann\'ees, pour essayer de discriminer les deux mod\`eles. Est-ce que les galaxies pr\'esentent le m\^eme taux d'instabilit\'es\,? Forment-elles \'egalement des barres, des spirales, qui sont le moteur de l'\'evolution et de la concentration de la masse\,? Les disques sous MOND sont plus instables, et forment des barres plus vite, mais lorsque les barres sont d\'etruites il est plus difficile de les former. Globalement, on peut obtenir la fr\'equence des barres observ\'ees dans les deux mod\`eles. Toutes les simulations d\'emontrent qu'il est possible de reproduire les observations avec MOND, m\^eme dans le cas d'interactions entre galaxies, comme le montre la Figure 5, illustrant la fusion entre deux galaxies, les ``Antennes''. 
\begin{figure}[t]
\centering{\includegraphics[width=5in]{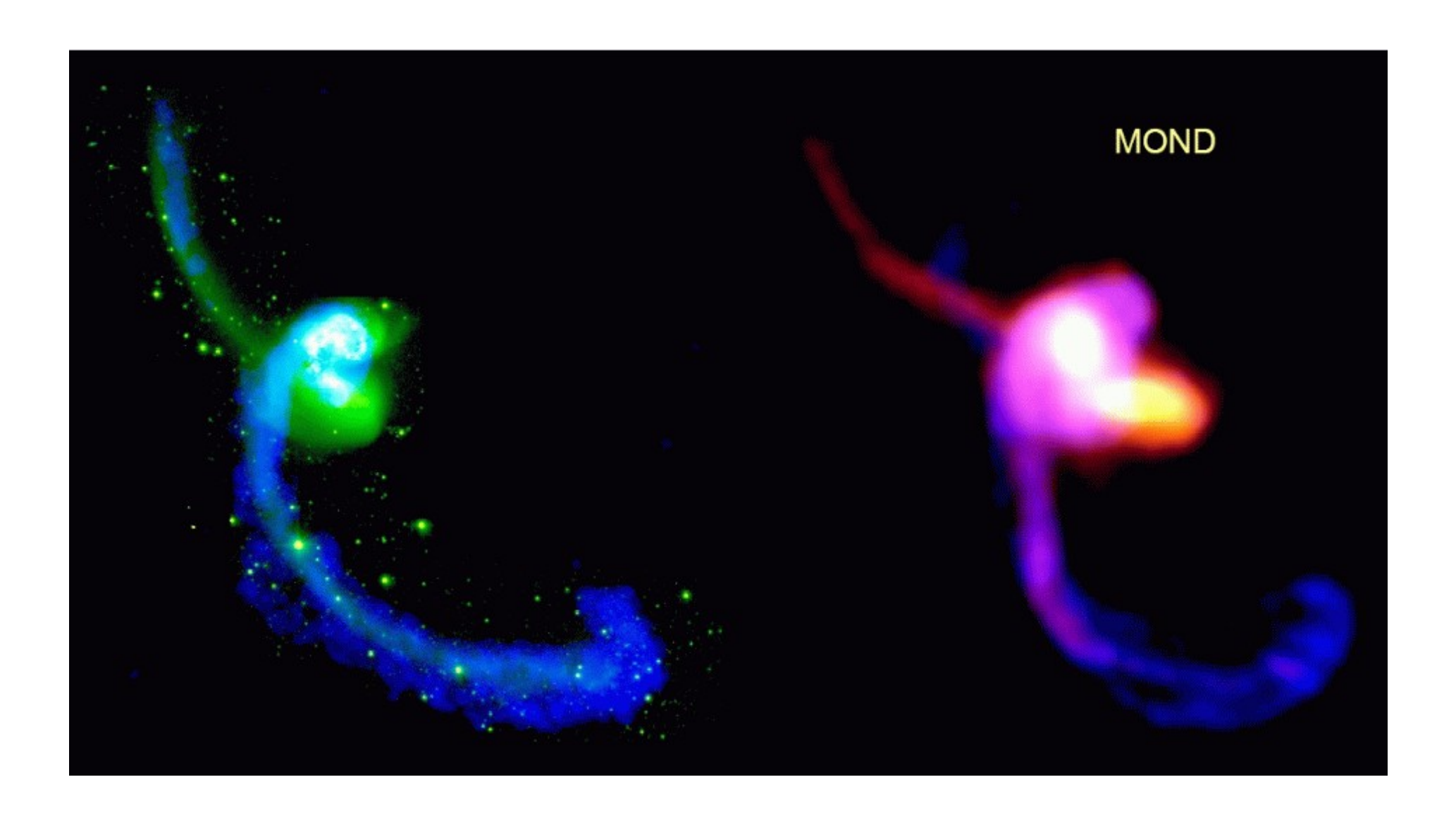}
\caption{Comparaison entre l'observation des ``Antennes'' \`a gauche, galaxies de masses semblables en interaction, et d\'eveloppant deux queues de mar\'ee, et les simulations dans le cadre de MOND, de deux galaxies de masse \'egale. D'apr\`es Tiret \& Combes (2008).}
\label{fig5}}
\end{figure} 

Si le formalisme MOND parvient \`a r\'esoudre les probl\`emes du mod\`ele CDM \`a l'\'echelle des galaxies, il rencontre en revanche ses propres probl\`emes \`a l'\'echelle des amas de galaxies. Dans cet environnement, l'acc\'el\'eration n'est pas toujours tr\`es faible par rapport \`a l'acc\'el\'eration critique $a_0$, et l'amplification de la gravit\'e n'est pas aussi forte qu'au niveau des galaxies. MOND rend compte d'une partie de la masse manquante, mais il en reste encore\,! Ce probl\`eme pourrait \^etre r\'esolu par la pr\'esence soit de neutrinos, particules tr\`es nombreuses dans l'Univers, et dont la masse a \'et\'e longtemps suppos\'ee nulle, mais serait de l'ordre de un \'electron-Volt en \'equivalent \'energie, soit de baryons noirs, ou mati\`ere ordinaire, faite d'atomes qui ne rayonnent pas, dans les amas. Ceci est encore possible, car la majeure partie des baryons n'ont pas encore \'et\'e identifi\'es. Nous savons, par les exp\'eriences rappel\'ees au d\'ebut, que les baryons constituent 4\% du contenu de l'Univers, et pourtant la mati\`ere visible (\'etoiles et gaz) ne recense que environ 10\% de ces baryons, soit 0,4\% du total. Les amas ne repr\'esentent qu'une tr\`es faible partie de la masse de l'Univers, et aucune contrainte n'existe alors sur la quantit\'e de baryons sombres qu'ils peuvent contenir.

\section{Gravit\'e modifi\'ee ou mati\`ere modifi\'ee\,?}

La formule MOND a montr\'e une performance \'etonnante pour les halos de
galaxies, mais a clairement un domaine d'application limit\'e. D'autre part ce
n'est qu'une ``recette'' qui ne rentre pas dans le cadre th\'eorique actuel. Et pourtant, elle semble dire quelque chose d'important sur le probl\`eme de la mati\`ere noire, et peut-\^etre m\^eme de l'\'energie noire car la constante $a_0$ se trouve \^etre du m\^eme ordre de grandeur que la valeur associ\'ee \`a la constante cosmologique $\Lambda$.

Une possibilit\'e est qu'un jour la formule MOND soit expliqu\'ee par une s\'erie de ph\'enom\`enes physiques qui n'ont pas encore \'et\'es pris en compte dans les simulations de CDM, tels que l'effet des supernov{\ae} sur la distribution de mati\`ere noire ou l'interaction de la mati\`ere noire avec les baryons. Cela semble n\'eanmoins improbable car on imagine difficilement pouvoir r\'esoudre ainsi le probl\`eme de la conspiration entre la mati\`ere noire et la mati\`ere visible. D'autre part, aucun m\'ecanisme convainquant n'a \'et\'e trouv\'e pour incorporer l'acc\'el\'eration $a_0$ dans le mod\`ele CDM.

La seconde solution est la gravit\'e modifi\'ee MOND sans mati\`ere noire, d\'ecrite plus haut. On cherche une modification de la relativit\'e g\'en\'erale d'Einstein de fa\c{c}on \`a reproduire MOND dans la limite non-relativiste. Il n'est pas facile de modifier les \'equations de la relativit\'e g\'en\'erale \`a cause de leur parfaite coh\'erence math\'ematique\,! On suppose donc que la relativit\'e g\'en\'erale est valable, mais on rajoute des nouveaux champs associ\'es \`a la gravitation, qui vont ob\'eir \`a de nouvelles \'equations, et vont aussi se coupler \`a la m\'etrique de l'espace-temps de la relativit\'e g\'en\'erale. Jacob Bekenstein (2004) a montr\'e qu'il est possible de construire une telle th\'eorie, bas\'ee sur des champs vectoriel et scalaires en plus du champ tensoriel habituel. Cette th\'eorie TeVeS (pour Tenseur-Vecteur-Scalaire) est le premier exemple de th\'eorie relativiste reproduisant MOND.

Une troisi\`eme alternative est propos\'ee par L. Blanchet et A. Le Tiec depuis deux ans\,: la mati\`ere modifi\'ee. On suppose que la mati\`ere noire est munie d'une propri\'et\'e qui la fait se comporter diff\'eremment de CDM, et permet d'expliquer la ph\'enom\'enologie de MOND, sans modifier la th\'eorie de la gravitation qui reste la relativit\'e g\'en\'erale. Cette approche a donc la couleur de MOND, mais ce n'est pas du MOND\,! 

\section{La mati\`ere noire dipolaire}

L'approche de mati\`ere modifi\'ee est bas\'ee sur une analogie remarquable avec la physique des dip\^oles \'electrostatiques. Dans un milieu di\'electrique, les atomes sont mod\'elis\'es par des dip\^oles \'electriques qui se polarisent en pr\'esence d'un champ \'electrique ext\'erieur. Le champ total est alors la somme du champ ext\'erieur et du champ dipolaire induit par la polarisation. Or MOND appara\^it exactement comme l'analogue gravitationnel d'un effet de polarisation (Blanchet 2007). Dans le cas gravitationnel la polarisation tend \`a augmenter l'intensit\'e du champ (au contraire du cas \'electrique o\`u le champ est ``\'ecrant\'e'' par les charges de polarisation), et c'est bien ce qu'il nous faut pour expliquer la mati\`ere noire.

Se pourrait-il que la mati\`ere noire soit constitu\'ee d'un milieu de dip\^oles gravitationnels\,? Si l'on poursuit l'analogie avec l'\'electrostatique, le dip\^ole gravitationnel devrait \^etre constitu\'e d'une masse n\'egative associ\'ee \`a une masse positive. La pr\'esence de masses n\'egatives conduit \`a une violation du principe d'\'equivalence, difficilement r\'econciliable avec le formalisme de la relativit\'e g\'en\'erale. N\'eanmoins, il est possible de d\'ecrire le dip\^ole gravitationnel de mani\`ere ``effective'', sans lui donner une interpr\'etation microscopique fondamentale. On peut alors construire un mod\`ele de mati\`ere noire dipolaire en relativit\'e g\'en\'erale (Blanchet \& Le Tiec 2008, 2009). Ce mod\`ele reproduit naturellement la ph\'enom\'enologie de MOND gr\^ace au m\'ecanisme de polarisation gravitationnelle, et d'autre part on trouve que la mati\`ere noire dipolaire se comporte comme CDM aux grandes \'echelles cosmologiques, et est donc en accord avec les fluctuations du fond cosmologique. Le mod\`ele est viable\,! Il permet d'unifier les deux facettes antagonistes de la mati\`ere noire, son aspect ``particulaire'' en cosmologie, et son aspect ``modification des lois'' dans les galaxies.

On le voit, les mod\`eles alternatifs ne manquent pas, et la question est ouverte\,: les exp\'eriences men\'ees par les t\'elescopes au sol et les satellites dans le futur, ainsi que peut-\^etre des d\'eveloppements th\'eoriques nouveaux, permettront de trancher.



\begin{thebibliography}{}
\bibitem{bek04} Bekenstein, J. 2004, Phys. Rev. D {\bf 70}, 083509.
\bibitem{BO7} Blanchet, L. 2007, Class. Quant. Grav. {\bf 24}, 3529.
\bibitem{BL08} Blanchet, L. \& Le Tiec, A. 2008, Phys. Rev. D {\bf 78}, 024031; ibid. 2009, Phys. Rev. D {\bf 80}, 023524.
\bibitem{clowe} Clowe, D., et al., Astrophys. J. {\bf 648}, L109 (2006).
\bibitem{diemand} Diemand, J., Kuhlen, M. \& Madau, P. 2007, Astrophys. J. {\bf 667}, 859.
\bibitem{milgrom} Milgrom, M. 1983, Astrophys. J. {\bf 270}, 365.
\bibitem{TC08} Tiret, O. \& Combes, F. 2008, Astron. Soc. Pac. Conf. {\bf 396}, 259.
\bibitem{springel} Springel, V., Frenk, C. \& White, S. 2006, Nature {\bf 440}, 1137.
\end{thebibliography}
\end{document}